\documentclass[aps,prd,noshowpacs,superscriptaddress,amsmath,amssymb,twocolumn,preprintnumbers]{revtex4}
\usepackage{amsmath,amssymb,lmodern}
\usepackage{bbold}
\usepackage{bm}
\usepackage{relsize}
\usepackage{braket}
\usepackage{bigints}
\usepackage{float}

\renewcommand{\Re}{\operatorname{Re}}

\renewcommand{\v}[0]{\bm}

\newcommand{\tat}{\textit}

\usepackage{graphicx}
\graphicspath{{figures/}}

\begin{document}
\title{Transverse spectrum of bremsstrahlung in finite condensed media}
\author{X.Feal}\email{xabier.feal@igfae.usc.es}
\author{R.A.Vazquez}\email{vazquez@igfae.usc.es}

\affiliation{ Departamento de F\'{\i}sica de Part\'{\i}culas 
              \& Instituto Galego de F\'\i sica de Altas Enerx\'\i as\\
              Universidade de Santiago de Compostela, 15782 Santiago, SPAIN}

\date{\today}

\begin{abstract}
A formalism is presented in which the radiation of photons off high energy
electrons during a multiple scattering process with finite condensed media can
be evaluated for a general interaction. We show that the arising
Landau-Pomeranchuk-Migdal suppression for finite size targets saturates at
some characteristic photon energy. Medium coherence effects in the photon
dispersion relation can be also considered leading to a dielectric suppression
or transition radiation effects in the soft part of the spectrum.  The main
results of our formulation are presented for a Debye screened interaction and
its well-known Fokker-Planck approximation, showing that for finite size
targets or for the angular distributions of the final particles the
differences between both scenarios cannot be reconciled into a single
redefinition of the medium transport parameter $(\hat{q})$.  Our predictions
are in very good agreement with the experimental data  collected at SLAC. 
\end{abstract}
\maketitle

\section{Introduction}

The Landau-Pomeranchuk-Migdal (LPM) suppression is a well known effect that
has been extensively studied. Interference phenomena in a multiple scattering
scenario was initially considered by Ter-Mikaelian as the mechanism regulating
the amount of scattering centers which can coherently emit as a single
bremsstrahlung source \cite{Termikaelian1953}.  A classical evaluation of this
effect for a semi-infinite medium was soon introduced by Landau and
Pomeranchuk \cite{Landau1953a,Landau1953b} and later completed by Migdal
\cite{Migdal1956} for the quantum case by means of a Boltzmann transport
equation for the electron. This calculation has shown that except for the spin
corrections for hard photons, the LPM suppression for an averaged target still
agrees with the expected classical behavior of the infrared
divergence. Further and more recent developments in various approaches have
been introduced since then by Blankenbecler and Drell
\cite{Blankenbecler1,Blankenbecler2,Blankenbecler3}, Zakharov
\cite{Zakharov2,Zakharov3,Zakharov1,Zakharov3a,Zakharov4,Zakharov5,Zakharov6},
the Baier-Dokshitzer-Mueller-Schiff-Peign\'e group (BDMPS) \cite{RBaier},
Baier and Katkov \cite{VBaier1,VBaier3,VBaier4} and Wiedemann and Gyulassy
\cite{Wiedemann1}, and extensive reviews were presented in
\cite{Klein,VBaier2}. We note, however, that all the existing calculations
were done in the Fokker-Planck approximation, which both in the Boltzmann
transport approach \cite{Migdal1956,RBaier} and in the path integral
formulation \cite{Zakharov2,Wiedemann1} lead to a Gaussian distribution of
momenta. In this approximation, then, the transport properties of the medium
have to be adequated to take into account the neglected large momentum tails
of the original Debye screened or Coulomb interactions. Few works, on the
other hand, considered the finite target case, which has always been
problematic and sometimes misunderstood lacking a general formulation. Also,
the angular distribution of the final particles has not been studied in
general \cite{Wiedemann1}. Taking into account these remarks, no result has
ever been given beyond the Fokker-Planck approximation, that also accounts for
the transverse photon and electron spectrum, and which includes in a natural
way the finiteness of the target.

We have developed a formalism and a Monte Carlo code which allows for the
computation of the bremsstrahlung spectrum of finite targets, arbitrary
interactions and with a full control of the kinematics of the process, so that
specific cuts on momenta of the final electron and photon can be applied.  In
section \ref{formalism} we will briefly explain the LPM effect, review the
formalism, and give several approximations for the calculation.  In
section \ref{results}, we will present and compare our results with the
experimental data of SLAC \cite{Anthony3}. Finally, we end in section
\ref{conclusions} with some conclusions.

\section{Formalism and calculation}
\label{formalism}

It has been predicted by Ter-Mikaelian \cite{Termikaelian1953} and Landau and
Pomeranchuk \cite{Landau1953a} that at high energies the Bethe-Heitler cross
section \cite{Bethe1934} stops being applicable to extended media. In order to
understand this phenomenon we start with the emission amplitude for a process
consisting in a collision with $(n)$ sources
\begin{align}
\mathcal{M}_{em}^{(n)} = -ie \int d^4y \medspace \bar{\psi}_f^{(n)}(y)
\gamma^\mu
A_\mu^\lambda(y)\psi_i^{(n)}(y)+\mathcal{O}(e^2)\label{amplitude_emission_m},
\end{align}
where $A_\mu^\lambda(y)=\mathcal{N}(k)\epsilon_\mu^\lambda e^{ik\cdot y}$ is a
free photon of momentum $k$ and polarization $\lambda$ and
$\mathcal{N}(k)=\sqrt{2\pi/\omega}$ its normalization, $\Psi_{i,f}^{(n)}(x)$
the incoming and outgoing electron wave functions under the external field of
the medium and $e=\sqrt{\alpha}$ the electron charge. Since in the $\omega\to
0$ limit the number of photons diverges, in virtue of the soft photon theorem
\cite{Weinberg1965}, the classical approximation holds \cite{bloch1937} and we
can replace
\begin{align}
J_{k}(x)=\bar{\Psi}_{f}^{(n)}(x)\gamma_k\Psi_{i}^{(n)}(x)\to J_{k}(x) \equiv
v_k(t)\medspace\delta^3(\v{x}-\v{x}(t)),
\end{align}
where $\v{v}(t)\equiv \dot{\v{x}}(t)$ is the electron velocity, yielding (see 
\cite{Jackson,Bell1958}
\begin{align}
\mathcal{M}_{em}^{(n)}=-ie\mathcal{N}(k)\int^{+\infty}_{-\infty} dt \medspace
\bigg(\frac{\v{k}}{\omega}\times \v{v}(t)\bigg)e^{i\omega
t-i\v{k}\cdot\v{x}(t)}, \label{amplitude_emission_m_classical}
\end{align}
and where we used $\v{\epsilon}^\lambda\cdot\v{k}=0$. One can consider the
integration over time as the point in which the photon is emitted.  This
observation becomes manifest by letting the electron describe a
discretized trajectory, with velocities $\v{v}_j$ for $j=1,\ldots,n_c+1$ and
piecewise path $\v{x}_j=\v{x}_{j-1}+\v{v}_{j-1}(t_j-t_{j-1})$, where $n_c$ is
the number of collisions. Equation \eqref{amplitude_emission_m_classical} then
produces
\begin{align}
\mathcal{M}_{em}^{(n)}=e\mathcal{N}(k)\frac{1}{\omega}\sum_{j=1}^{n_c}\v{\delta}_j 
\medspace  
e^{i\varphi_j},\label{amplitude_emission_m_classical_discretized}
\end{align}
where we find a superposition of $n_c$ single Bethe-Heitler like amplitudes
\cite{Bethe1934} of the form
\begin{align}
\v{\delta}_j \equiv \v{k} \times
\bigg(\frac{\v{v}_{j+1}}{\omega-\v{k}\cdot\v{v}_{j+1}}-\frac{\v{v}_{j}}
{\omega-\v{k} \cdot\v{v}_{j}}\bigg),\label{single_bethe_heitler}
\end{align}
interfering with a phase $\varphi_j\equiv \omega t_j-\v{k}\cdot\v{x}_j$. The
evaluation of the square of \eqref{amplitude_emission_m_classical_discretized}
leads to a total emission intensity between the photon solid angle
$\Omega_k$ and $\Omega_k+d\Omega_k$ given by
\begin{align}
\omega \frac{dI}{d\omega d\Omega_k} = \frac{e^2}{(2\pi)^2} \left(\sum_{j=1}^{n_c}
\v{\delta}_j^2+2\Re\sum_{j=1}^{n_c}\sum_{i=1}^{j-1} \v{\delta}_j\cdot \v{\delta}_i
e^{i\varphi_{i}^{j}}\right),\label{classicalintensity}
\end{align}
where we have split the sum in a diagonal and a non-diagonal contribution.
The interfering behavior of the above sum is governed by the phase change
between two arbitrary collisions or emission elements
\begin{align}
\varphi_{i}^j&\equiv \varphi_j-\varphi_i = k_\mu (x_j^\mu-x_i^\mu) =
\int^{z_j}_{z_i} dz \frac{k_\mu p^\mu(z)}{p_0}\nonumber\\ &=
\omega(1-\beta)\int_{z_i}^{z_j} dz +\omega
\int_{z_i}^{z_j}dz \medspace\frac{\delta \v{p}^2(z)}{2\beta p_0^2},
\label{phase_definition}
\end{align}
where $p_0$ is the initial electron energy, $\delta\v{p}(z)$ is the
accumulated momentum change of the electron at $z$ with respect to the photon
direction and $\beta=|\v{v}|=\sqrt{1-m_e^2/p_0^2}$ the electron velocity.
This phase can be made maximal for large emission angles and/or photon
frequencies, provided that $\varphi_{i}^{i+1}\gg 1$ for any two consecutive
collisions. In that case the non-diagonal sum in \eqref{classicalintensity}
cancels and we are left with a totally incoherent superposition of $(n_c)$
single Bethe-Heitler intensities, with a maximal intensity of
\begin{align}
\omega \frac{dI_{sup}}{d\omega d\Omega_k} = \frac{e^2}{(2\pi)^2}
\sum_{j=1}^{n_c}\v{\delta}_j^2.\label{incoherent_plateau_classical}
\end{align}
In this regime emission decouples and all the scatterings can be considered to
be independently emitting.  In the opposite case, when the emission angle
and/or photon energy are small so that the phase vanishes, the internal
structure of the scattering is irrelevant.  This observation becomes manifest by
setting $\varphi_j=0$ in \eqref{amplitude_emission_m_classical_discretized},
so we are left with the first and last terms only and intensity acquires the
minimum value
\begin{align}
\omega \frac{dI_{inf}}{d\omega d\Omega_k} = \frac{e^2}{(2\pi)^2}
\left|\sum_{j=1}^{n_c}\v{\delta}_j\right|^2,\label{coherent_plateau_classical}
\end{align}
which can be interpreted as a Bethe-Heitler intensity with a final velocity
$\v{v}_{n_c+1}$ due to the coherent deflection with all the medium centers.
In this regime the entire medium acts as a single independent emission
element. This behavior is a consequence of the well known soft photon theorem
\cite{Weinberg1965,Weinberg}, although in the LPM literature it is known as
the Ternovskii-Shul'ga-Fomin emission \cite{Ternovskii1,ShulgaFomin}. The suppression from the
superior (incoherent) plateau of radiation
\eqref{incoherent_plateau_classical} to the inferior (coherent) plateau
\eqref{coherent_plateau_classical} is known as the LPM effect for mediums of
arbitrary size. Notice that in the infinite medium limit ($n_c\gg 1$) the
coherent plateau can be neglected, since the soft photon theorem is not
observed and then the suppression is infinite.
\begin{figure}
\includegraphics[scale=0.25]{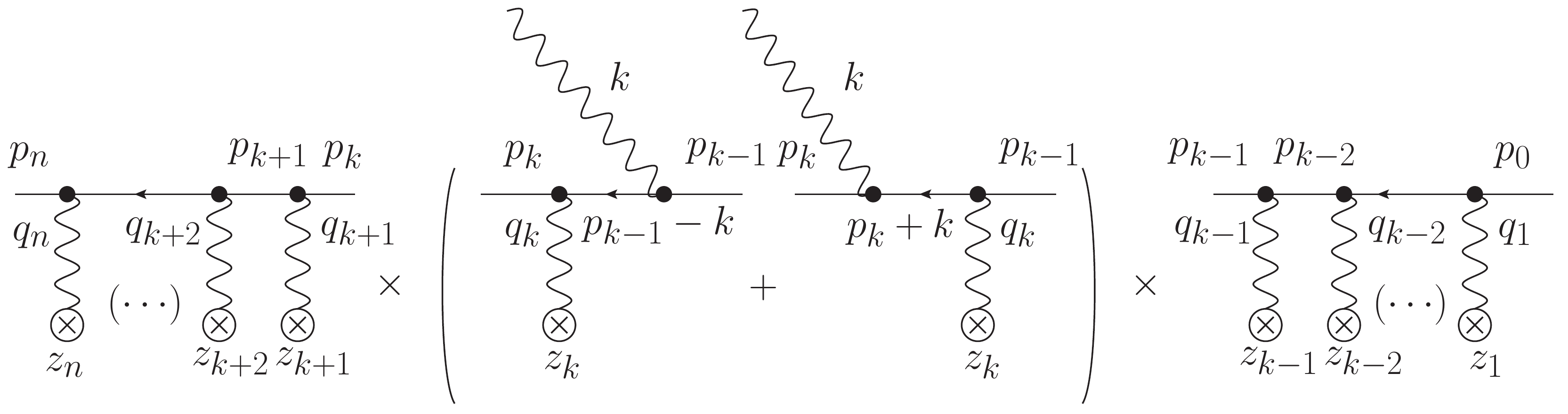}
\caption{Diagrammatic representation of the single emission elements appearing
  in a discretization in the variable $z$, of the amplitude
  $\mathcal{M}_{em}^{(n)}$ given at Equation
  \eqref{amplitude_emission_m_quantum_discretized}.}
\label{fig:image_diagram}
\end{figure}
The above classical arguments can be made quantitative and hold also for a
quantum evaluation of the amplitude. By Fourier transforming electron states
$\Psi_{i,f}(x)$ to the momentum space, we can write for the quantum amplitude
\eqref{amplitude_emission_m}
\begin{align}
&\mathcal{M}_{em}^{(n)}=e\mathcal{N}(k) \int
  \frac{d^3\v{p}(z)}{(2\pi)^3}\int^l_0 dz \exp\left(i \frac{k_\mu
    p^\mu(z)}{p_0}\right)
  \frac{d}{dz}\times\nonumber\\ &\left\{S_{s_ns}^{el}(p(l),p(z);l,z)
  \frac{f_{ss'}^\lambda(z)}{k_\mu
    p^\mu(z)}S_{s's_0}^{el}(p(z)+k,p(0);z,0)\right\},
\label{amplitude_emission_m_quantum_discretized}
\end{align} 
where we used the shorthand notation
\begin{align}
f_{ss'}^\lambda(z)\equiv \epsilon_\mu^\lambda(k) p_0
\sqrt{\frac{m_e}{p_0-\omega}}\bar{u}_{s}(p)\gamma^\mu
u_{s'}(p+k)\sqrt{\frac{m_e}{p_0}}.\label{spin_vertex}
\end{align}
Here $S_{s_2s_1}^{el}(p_2,p_1;l_2,l_1)$ stands for the beyond eikonal
evaluation of the elastic amplitudes for an electron to go from momentum $p_1$
to $p_2$ and from spin $s_1$ to $s_2$ due the amount of matter between $l_1$
and $l_2$, thus given by
\begin{align}
&S_{s_ns_0}^{(n)}(p_n,p_0;z_n,z_1) = 2\pi\delta(p_n^0-p_0^0)
\delta_{s_0}^{s_n}\beta 
\left(\prod_{i=i}^{n-1}\int\frac{d^2\v{p}^t_i}{(2\pi)^2}\right)\nonumber\\
&\times\left(\prod_{i=1}^{n}\int
d^2\v{x}^t_ie^{-i\v{q}_i\cdot\v{x}_i}\exp\left[-i\frac{g}{\beta}\sum_{j=1}^{n(z_i)}
 \chi_0^{(1)}(\v{x}-\v{r}_j)\right]\right),
\label{orderedsmatrix_discretized_b}
\end{align}
where we discretized the medium and thus $\v{q}_i\equiv \v{p}_i-\v{p}_{i-1}$
is the 3-momentum transfer at the layer $(i)$ of $n(z_i)$ scattering centers.
The external field characterizing the medium is given by $(n)$ single Debye
static sources with screening $\mu_d\simeq \alpha m_e Z^{1/3}$, coupled with
strength $g=Z\alpha$ to the electron, of the form
\begin{align}
\chi_0^{(1)}(\v{x})\equiv \int_{-\infty}^{+\infty} ds \medspace
A_0^{(1)}(\v{x}), \medspace
A_0^{(1)}(\v{x})=\frac{Z\alpha}{|\v{x}|}e^{-\mu_d|\v{x}|}.
\label{external_field}
\end{align}
The amplitude \eqref{amplitude_emission_m_quantum_discretized}, which
corresponds to a sum of the single emission elements shown in
Fig.(\ref{fig:image_diagram}), can be squared and averaged over medium
configurations of infinite transverse size $R\to \infty$ in a finite length
$l$, summed over final states, and averaged over initial states, leading to an
intensity of emission in the photon solid angle $\Omega_k$ and
$\Omega_k+d\Omega_k$ and per unit of medium transverse size and unit time of
\begin{align}
\omega &\frac{dI}{d\omega d\Omega_k} = \left(\frac{e}{2\pi}\right)^2
\left(\prod_{k=1}^{n}\int \frac{d^3\v{p}_j}{(2\pi)^3}\right)
\left(\prod_{k=1}^{n}\phi(\delta\v{p}_k,\delta
z)\right)\nonumber\\ &\times\left(h^n(y)\left|\sum_{j=1}^n \v\delta_j^n
e^{i\varphi_{0}^j} \right|^2+h^s(y)\left|\sum_{j=1}^n
\delta_j^se^{i\varphi_{0}^j} \right|^2\right),
\label{quantum_intensity}
\end{align}
where the spin non-flip currents $\v\delta_j^n$ are given by
\eqref{single_bethe_heitler} and the spin flip currents are given by
\begin{align}
\delta_j^s\equiv
\frac{1}{1-\beta_k\hat{\v{k}}\cdot\v{v}_{j+1}}-\frac{1}{1-\beta_k\hat{\v{k}}
\cdot\v{v}_{j}}.
\label{single_bethe_heitler_quantum}
\end{align}
Here we have introduced explicitely a medium with a refractive index
$1/\beta_k$ and $\beta_k$ is the photon velocity.  The functions $h^n(y)$
and $h^s(y)$ are the diagonal and non-diagonal sum in spins and helicities of
the squared emission vertex \eqref{spin_vertex}, given by
\begin{align}
h^n(y)=\frac{1}{2}(1+(1-y)^2),\medspace h^s(y)=\frac{1}{2}y^2,
\end{align}
and $y=\omega/p_0$ is the fraction of energy carried by the photon. They
produce two contributions of the same order, the last one, however, only
noticeable when $y\approx 1$ due to $h^s(y)$.  In what follows we will neglect
this contribution by assuming that $y\ll 1$. Within the same approximation we
will assume also that the electron 4-momentum change in the emission vertex is
negligible and $\beta=1$ unless otherwise required.  The local elastic weights
arising in the averaging of the square of \eqref{orderedsmatrix_discretized_b}
are given by
\begin{align}
\phi(\delta\v{p},\delta z)&= \exp\bigg(-n_0(z)\delta z \sigma_t^{(1)}
\bigg)(2\pi)^3\delta^3(\delta\v{p})\nonumber\\ &+2\pi\delta(\delta
p_0)\beta\Sigma_2(\delta\v{p},\delta z),
\label{elastic_weight}
\end{align}
where we can define the no collision probability $\exp(-n(z)\delta z
\sigma_t^{(1)})$ in the layer of length $\delta z$ and density $n_0(z)$ times
the forward distribution $\delta^3(\delta\v{p})$, and the collisional
distribution $\Sigma_2(\v{q},\delta z)$ after an incoherent scattering with
the centers in $\delta z$
\begin{align}
\Sigma_2(\v{q},\delta z) &\equiv \int d^2\v{x}
e^{-i\v{q}\cdot\v{x}}\exp\bigg(-n_0(z)\delta z
\sigma_t^{(1)}\bigg)\label{moliere_solution}\\ &
\times\bigg(\exp\bigg(n_0(z)\delta z\sigma(\v{x})\bigg)-1\bigg),
\nonumber
\end{align}
which satisfies a Moliere's equation with boundary condition
$\Sigma_2(\v{q},0)=0$. The required single elastic cross sections at
\eqref{elastic_weight} and \eqref{moliere_solution} can be shown to satisfy
$\sigma_t^{(1)}\equiv\sigma(\v{0})$ where, at leading order in $Z\alpha$ using
\eqref{external_field}
\begin{align}
\sigma(\v{x}) \equiv \frac{4\pi
  (Z\alpha)^2}{\beta^2\mu_d^2}\mu_d|\v{x}|K_1(\mu_d|\v{x}|)+
\mathcal{O}(Z\alpha)^3.
\end{align}
\begin{figure}
\includegraphics[scale=0.6]{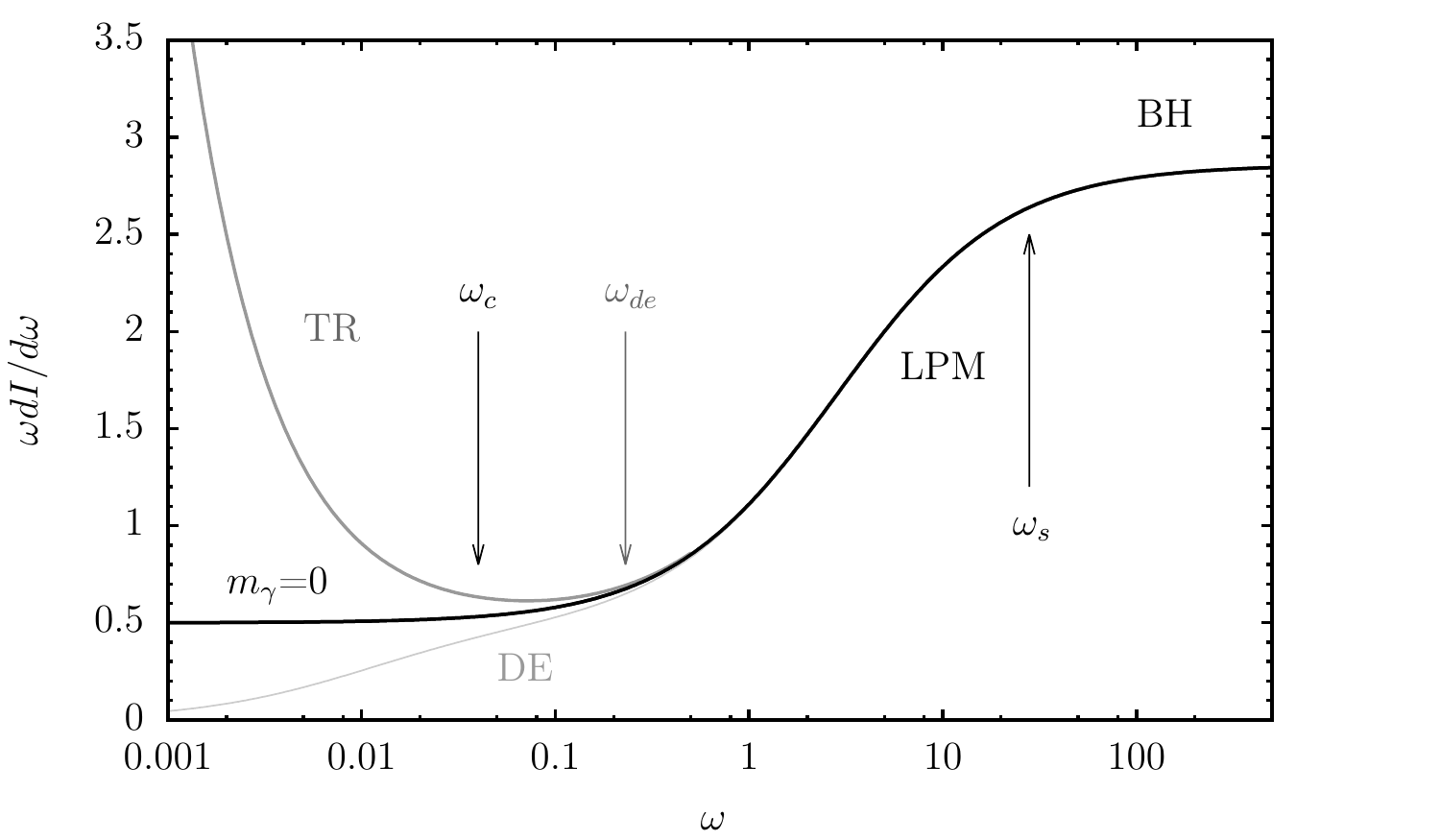}
\caption{Schematic representation of the bremsstrahlung regimes for several
  scenarios. Totally incoherent Bethe-Heitler superposition (BH),
  Landau-Pomeranchuk-Migdal suppression (LPM), totally coherent Bethe-Heitler
  superposition ($m_\gamma$=0), dielectric suppression (DE) and transition
  radiation (TR). See text for definition of the characteristics frequencies
  $\omega_c, \omega_s, \omega_{de}$}
\label{fig:image0}
\end{figure}
We assume from here onwards a constant density $n_0(z)\equiv n_0$ so from
\eqref{elastic_weight} the electron mean free path is read $\lambda\equiv
1/n_0\sigma_t^{(1)}$. In this elastic propagation the electron acquires a
squared momentum transfer additive with the traveled length $l$. Indeed, from
\eqref{elastic_weight} we find
\begin{align}
\frac{\partial}{\partial l} \left\langle \delta\v{p}^2(l)\right\rangle =
n_0\sigma_t^{(1)}\left\langle \delta\v{p}^2(\delta l)\right\rangle \equiv
2\hat{q},
\label{transport_parameter_definition}
\end{align}
where we defined the transport parameter $\hat{q}$. The momentum transfer in a
single collision $\delta l\lesssim \lambda$ is given, using
\eqref{elastic_weight}, by
\begin{align}
\left\langle\delta\v{p}^2(\delta l)\right\rangle =\mu_d^2
\left(2\log\left(\frac{2p_0}{\mu_d}\right)-1\right)=\mu_d^2\eta, 
\label{single_momentum_transfer}
\end{align}
where the correction $\eta$ to $\mu_d^2$ takes into account the long tail of
the Debye interaction \eqref{external_field} and a maximum momentum transfer
of $|\delta\v{p}|=2p_0$ is allowed in a single collision. High momentum
changes are suppressed at high energies, however, due to the functions
\eqref{single_bethe_heitler} and \eqref{single_bethe_heitler_quantum} in
\eqref{quantum_intensity}.  We have checked that a maximum momentum transfer
of $|\delta\v{p}|\simeq 2.5 m_e$ matches the single emission and then we write
for $\eta$ in \eqref{single_momentum_transfer} instead
\begin{align}
\eta=
\left(2\log\left(\frac{2.5m_e}{\mu_d}\right)-1\right)=
\left(2\log\left(\frac{2.5}{\alpha Z^{1/3}}\right)-1\right).
\label{effective_single_momentum_tranfer}
\end{align}
This effective momentum transfer under bremsstrahlung agrees with Bethe's
\cite{Bethe1934} estimation $\eta = 2\log(183/Z^{1/3})$ within less than
$3\%$ deviation in the range $Z=(1,100)$. Using
\eqref{transport_parameter_definition}, \eqref{single_momentum_transfer} and
\eqref{effective_single_momentum_tranfer} then $\hat{q}=(\eta/2) \times
n_0\sigma_t^{(1)} \mu_d^2 $ and the Fokker-Planck approximation for
\eqref{moliere_solution} reads
\begin{align}
\Sigma_{2}^G(\v{q},\delta z) \equiv \frac{2\pi}{\hat{q}\delta
  z}\exp\left(-\frac{\v{q}^2}{2\hat{q}\delta z}\right).
\label{moliere_solution_gaussian}
\end{align}
The above relations hold, however, for the single scattering regime $\delta
z\le \delta l$, so they can be used only to fix $\eta$ and thus $\hat{q}$ in
the incoherent plateau. For the coherent plateau a medium-length dependent fit
for $\eta$ has to be employed. Correspondingly, a single Fokker-Planck
approximation can not fit both the upper and lower ends of the bremsstrahlung
spectrum unless the medium length is very large, in which case the lower
plateau occurs at very low frequencies and can be neglected.

Before evaluating the expression \eqref{quantum_intensity}, we will derive an
heuristic formula for finite size targets to qualitatively understand
the interference phenomena. The coherence length defined by the phase
modulates the amount of scatterers which can be considered a single and
independent unit of emission in the squared amplitude. We then define the
length $\delta l=z_j-z_i$ in which the phase becomes larger than unity, which
using \eqref{phase_definition} becomes
\begin{align}
\varphi^{j}_i  \simeq
\frac{\omega}{2p_0^2} \left(m_e^2\delta l+\hat{q}(\delta l) ^2\right)\equiv 1,
\end{align}
then we get
\begin{align}
\delta l(\omega) \equiv
\frac{m_e^2}{2\hat{q}}\left(\sqrt{1+\frac{8\hat{q}p_0^2}{m_e^4\omega}}-1\right).
\label{coherence_length}
\end{align}
We also define the frequency $\omega_c$ at which the coherence length becomes
equal to $l$ thus $\omega_c\simeq p_0^2/(m_e^2l+\hat{q}l^2)$, and the
frequency $\omega_s$ in which the coherence length becomes equal to a mean
free path $\lambda$, $\omega_s\simeq
p_0^2/(m_e^2\lambda+\hat{q} l \lambda/2)$. Since the medium is finite we further
impose to \eqref{coherence_length} $\delta l(\omega)=l$ for
$\omega>\omega_c$. In the coherence length $\delta l(\omega)$ the internal
scattering structure is irrelevant since the phase can be neglected, and the
centers in $\delta l(\omega)$ act coherently like a single scattering source
with the total equivalent charge in $\delta l(\omega)$.  Since in the entire
medium $l$ there are $l/\delta l(\omega)$ coherence lengths, then we write the
incoherent sum
\begin{align}
\omega \frac{dI}{d\omega}(l) &= \frac{l}{\delta l(\omega)}e^2\int
\frac{d\Omega_k}{(2\pi)^2} \int
\frac{d^3\delta\v{p}}{(2\pi)^3}\nonumber\\ 
&\times\left(h^n(y)|\v{\delta}_1^n|^2+h^s(y)|\delta_1^s|^2\right)
\phi(\delta\v{p},\delta l(\omega)).
\end{align}
By integrating in the photon solid angle $\Omega_k$ and using
\eqref{elastic_weight} we find
\begin{align}
\omega \frac{dI(l)}{d\omega}=\frac{l}{\delta l(\omega)}\frac{e^2}{\pi^2}
\int_0^\pi d\theta \sin(\theta) F(\theta)\Sigma_2(\delta\v{p},\delta
l(\omega))
\label{heuristic_formula},
\end{align}
where the electron momentum change is $|\delta\v{p}|=2p_0\beta\sin(\theta/2)$
and
\begin{align}
F(\theta)=\bigg[&\frac{1-\beta^2\cos\theta}{2\beta\sin(\theta/2)
\sqrt{1-\beta^2\cos^2(\theta/2)}}\\
&\log\bigg[\frac{\sqrt{1-\beta^2
\cos^2(\theta/2)}+\beta\sin(\theta/2)}{\sqrt{1-\beta^2
\cos^2(\theta/2)}-\beta\sin(\theta/2)}\bigg]-1\bigg].\nonumber
\end{align}
This last integral \eqref{heuristic_formula} can be numerically evaluated and
the resulting intensity is exact for $\omega\gg \omega_s$ and $\omega\ll
\omega_c$.  A simple interpolation formula in the Fokker-Planck approximation
can be obtained from \eqref{heuristic_formula} by integrating its two
asymptotic values, i.e. $\delta l(\omega)\gg 1$ and $\delta l(\omega)\ll 1$,
and then interpolating both regions. One finds
\begin{figure}
\includegraphics[scale=0.6]{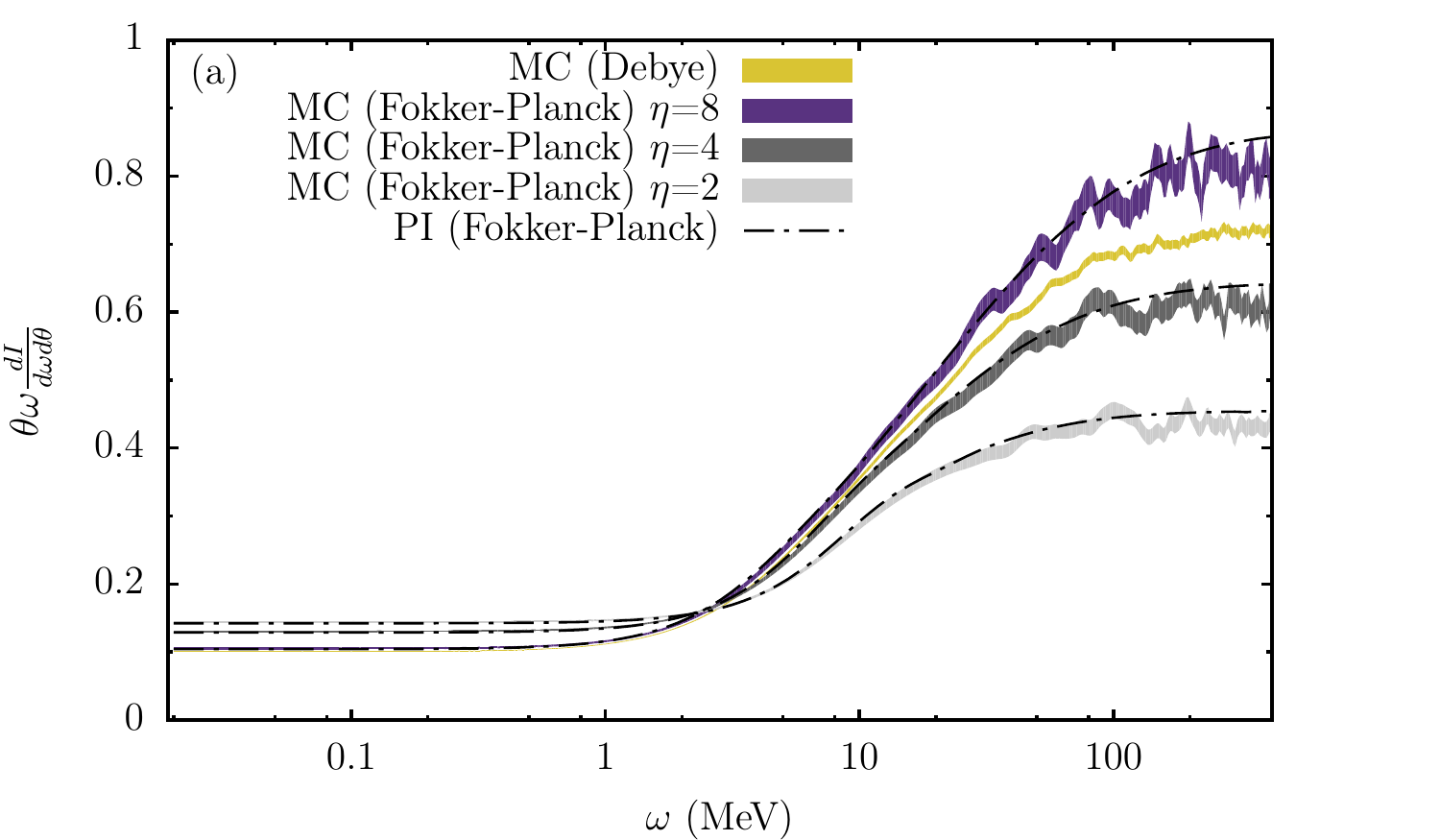}
\includegraphics[scale=0.6]{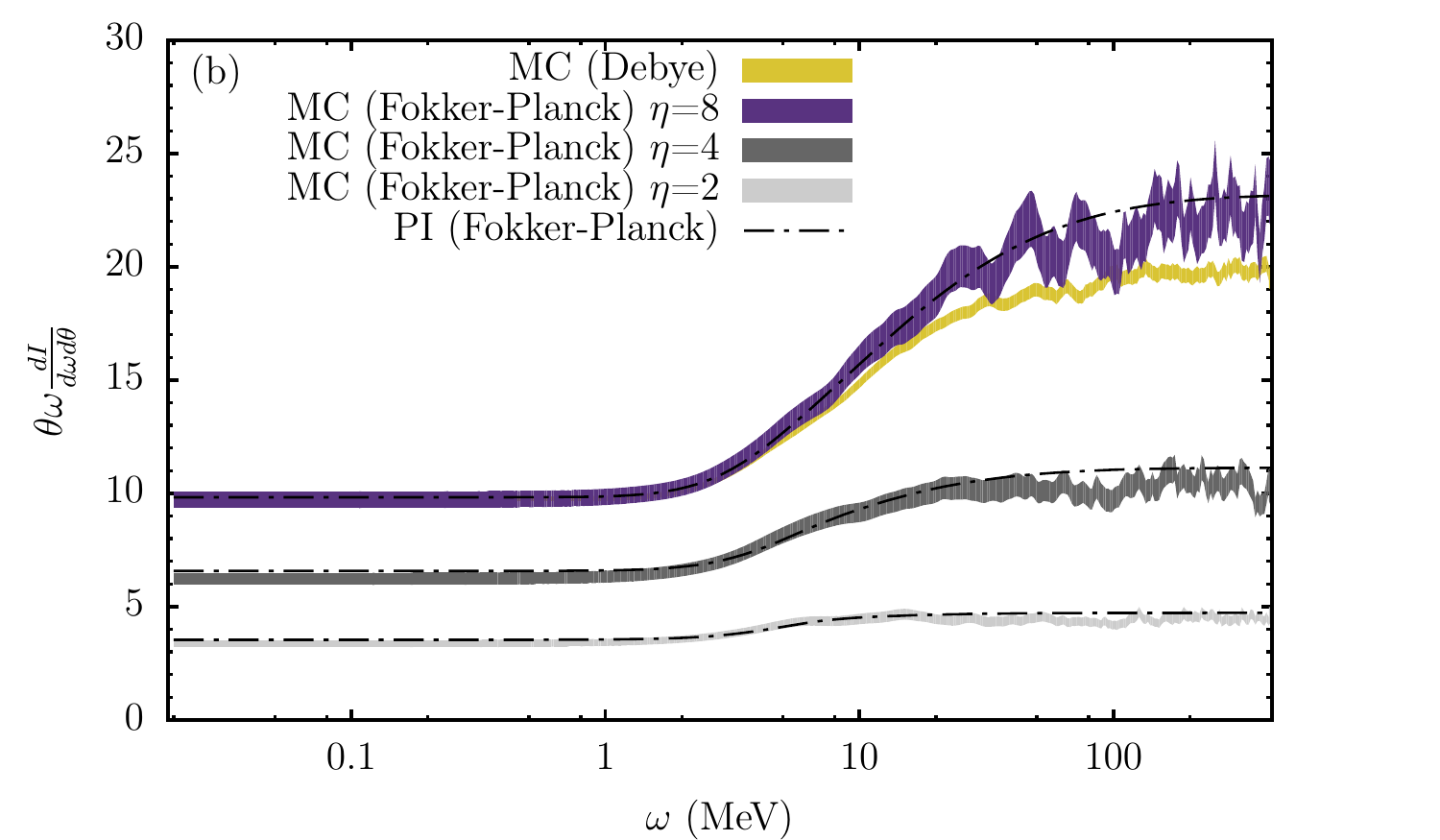}
\includegraphics[scale=0.6]{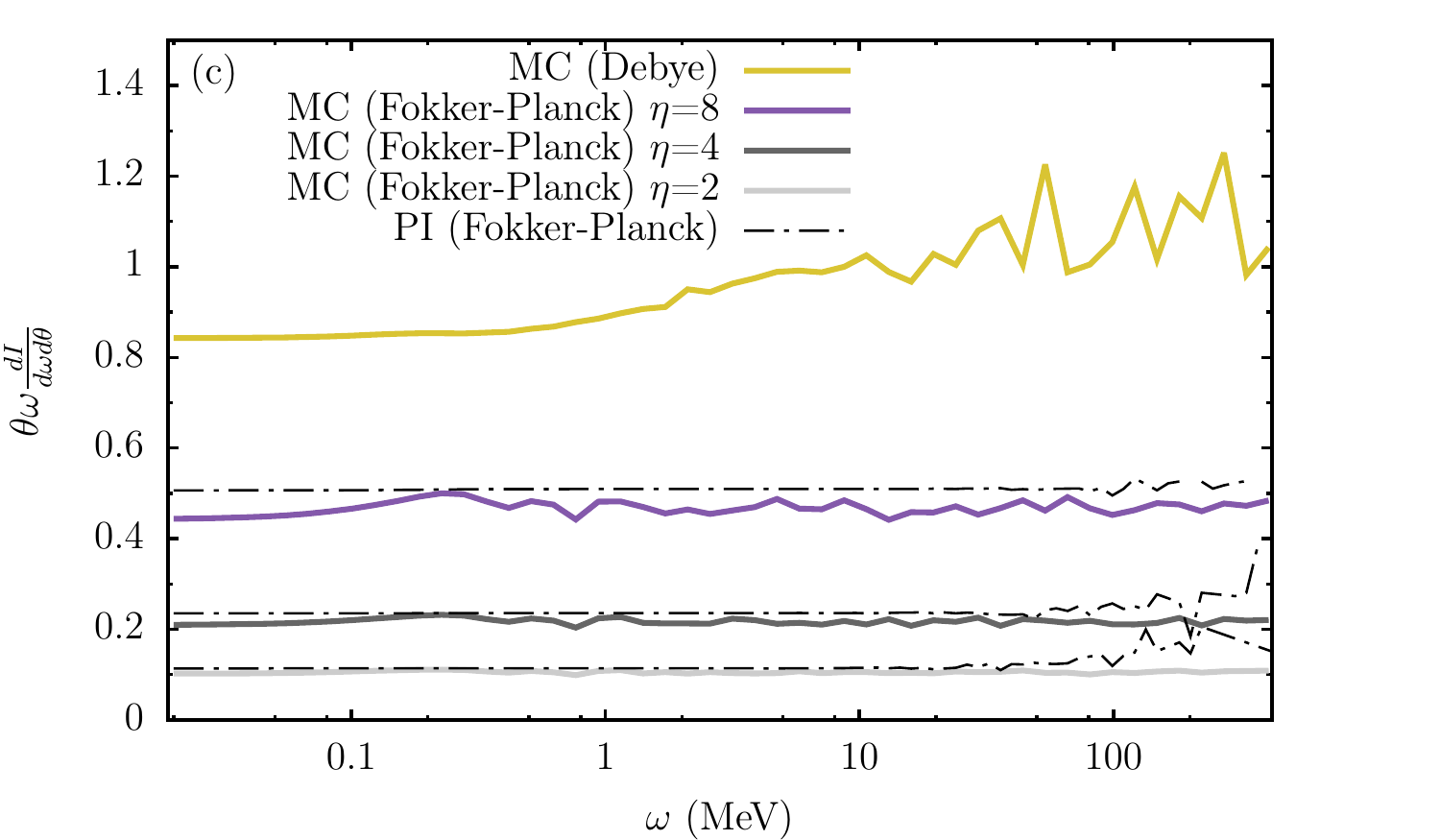}
\caption{Differential spectrum of photons emitted by an electron of $p_0 = 8$
  GeV traversing a sheet of gold of $l=0.023$ mm for photon angles
  $\theta_k=0.01/\gamma$ (a), $\theta_k=2/\gamma$ (b) and
  $\theta_k=10/\gamma$ (c), where $\gamma\equiv p_0/m_e$, in the Monte Carlo
  (MC) evaluations of \eqref{quantum_intensity} in the Debye interaction
  (yellow), in the Fokker-Planck approximation with $\eta=8$ (purple),
  $\eta=4$ (dark grey) and $\eta=2$ (light grey), together with the
  corresponding path integral (PI) limits of \eqref{quantum_intensity} in the
  Fokker-Planck approximation (dot-dashed lines). Bands show the statistical
  uncertainty of the Monte Carlo.}
\label{fig:image1}
\end{figure}

\begin{align}
\omega \frac{dI(l)}{d\omega}=\frac{l}{\delta l(\omega)}
\frac{2e^2}{\pi}\frac{1+n_m(\omega)}{3A+n_m(\omega)}
\log\bigg(1+An_m(\omega)\bigg),
\end{align}
where $n_m(\omega)\simeq 2\hat{q}\delta l(\omega)/m_e^2$ is a measure of the
accumulated transverse momentum in a coherence length and $A=e^{-(1+\gamma)}$ with
$\gamma$ Euler's constant. For completeness we also write Migdal $y\ll 1$
prediction \cite{Migdal1956} for semi-infinite mediums
\begin{align}
\omega
&\frac{dI(l)}{d\omega}=l\frac{2e^2}{\pi}\sqrt{\frac{\hat{q}\omega}{2p_0^2}}
\int_0^\infty dz \medspace\exp
\left(-\frac{z}{\sqrt{2}s}\right)\label{migdal_prediction_noapproximant}\\ &\times\left(\sin\left(\frac{z}
     {\sqrt{2}s}\right)+\cos\left(\frac{z}{\sqrt{2}s}\right)\right)
     \left(\frac{1}{z^2}-\frac{1}{\sinh^2(z)}\right),\nonumber
\end{align}
where $s\equiv (2p_0/m_e^2)\sqrt{\hat{q}/\omega}$ and an useful approximant
within less than a $1\%$ of deviation is given by
\begin{align}
\omega
\frac{dI(l)}{d\omega}=\frac{2e^2}{3\pi}\frac{2\hat{q}l}{m_e^2}\frac{1-1.52 s^4+5.8s^5}{1+2.44s^5+2.73s^6}.
\label{migdal_prediction}
\end{align}
Migdal result \eqref{migdal_prediction_noapproximant} is exactly recovered
within our formalism as a restriction in the domain of integration in $z$
of the emission amplitude \eqref{amplitude_emission_m} squared, as follows
\begin{align}
\left|\mathcal{M}_{em}^{(n)}\right|^2 = e^2&\int_{-\infty}^{+\infty}dz
\int_{-\infty}^{+\infty}dz'\\
\to e^2&\left\{\int^{+\infty}_0
dz\int^{+\infty}_0 dz'
+\int^{0}_{-\infty}dz\int^{0}_{-\infty}dz'\right\}\nonumber.
\end{align}  
The general behavior of these results can be summarized in
Fig.(\ref{fig:image0}) where the photon intensity is pictured as a function of
the photon frequency. Above the saturation frequency $\omega_s$ the photon
resolves all the internal structure of the scattering, the medium emits as a
total incoherent sum of $n_c=l/\lambda$ Bethe-Heitler intensities, where $n_c$
is the average number of collisions. In this regime, the photon intensity
scales with $l$. Notice that total suppression can occur provided $\omega_s$
becomes larger than $p_0$, which causes that electrons with energies below
$p_0^{lpm}=m^2_e/n_0\sigma_t^{(1)}$ experiment the bremsstrahlung suppression
in all their spectrum. For smaller frequencies the number of independent
emitters, using \eqref{coherence_length}, decreases with $\sqrt{\omega}$
whereas the charge of each element logarithmically grows with
$\log(1/\sqrt{\omega})$. This suppression stops at $\omega_c$, where the
coherence length \eqref{coherence_length} acquires the maximum value $l$, the
medium emits as a single entity and intensity saturates to Weinberg's soft
photon theorem \cite{Weinberg1965}.
\begin{figure}
\includegraphics[scale=0.6]{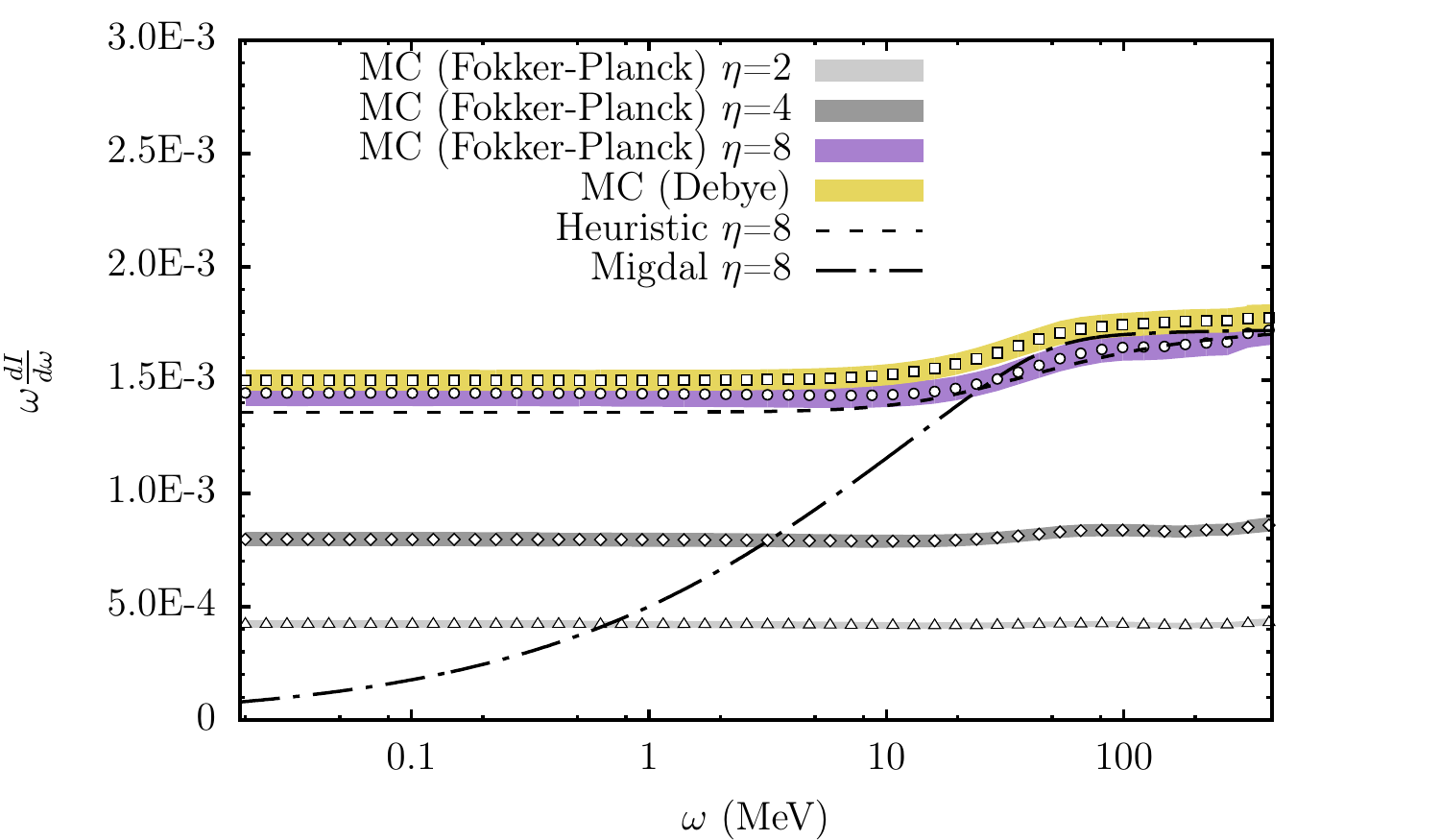}
\includegraphics[scale=0.6]{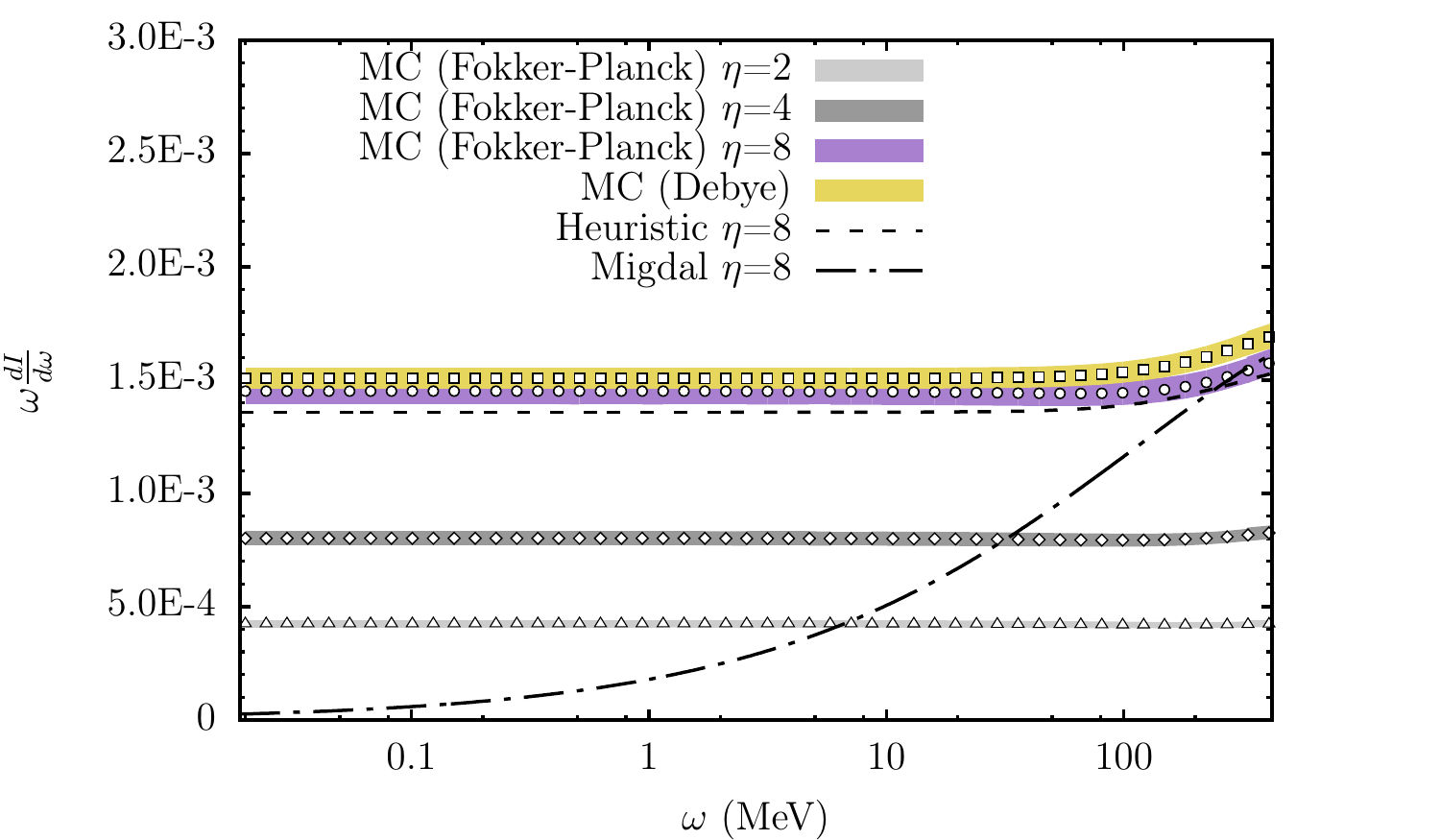}
\caption{Intensity of photons emitted by an electron of $p_0=8$ GeV (top) and
  $p_0=25$ GeV (bottom) after traversing a sheet of gold of length
  $l=0.0038$ mm in the Monte Carlo (MC) evaluation of \eqref{quantum_intensity}
  in the Debye interaction (squares), the Fokker-Planck approximation with
  $\eta=8$ (circles), $\eta=4$ (diamonds) and $\eta=2$ (triangles). Also shown
  our heuristic formula \eqref{heuristic_formula} and the Migdal prediction
  \eqref{migdal_prediction}.}
\label{fig:image2}
\end{figure}
The presence of a medium modifies the photon dispersion relation and
substantially changes this picture in the soft limit.  For the energies
considered here the photon has velocity given by
\begin{align}
\beta^2(k) = 1- \frac{\omega_p^2}{\omega^2},
\end{align}
where $\omega_p^2 \simeq 4\pi Z \alpha n_0/m_e \equiv m_\gamma^2$ is the
plasma frequency, which can be interpreted as a photon mass $m_\gamma$. This
scenario induces an additional source of suppression due to the fact that the
wavelength of a photon of frequency $\omega$ is now larger than in the vacuum
case and thus
\begin{align}
k_\mu p^\mu(\omega_p) \simeq k_\mu
p^\mu(0)+\frac{m_\gamma^2}{2\omega}.\label{plasma_frequency_phase}
\end{align}
This extra term further suppresses the coherent plateau at $\omega <
\omega_{de}$, where $k_\mu p^\mu(0)\equiv m_\gamma^2/2\omega_{de}$,
i.e. $\omega^2_{de}=\omega_p^2l\omega_c$, since the denominators of
\eqref{single_bethe_heitler} and \eqref{single_bethe_heitler_quantum}, defined
by \eqref{plasma_frequency_phase}, grow for smaller frequencies.  This
suppression is called the dielectric effect and holds for a totally
homogeneous space or infinite medium. However, if the electron passes through
vacuum to a medium and then again to vacuum, or in general through structured
targets where density cannot be considered constant, then the definition
\eqref{plasma_frequency_phase} becomes local for each photon emission point.
The photon emitted at the last leg then satisfies $m_\gamma=0$, whereas the
first leg photon satisfies $m_\gamma\neq 0$, thus it can be shown that an
interference destroys the dielectric suppression in the coherence plateau,
dramatically enhancing the intensity for $\omega<\omega_{de}$. This is called
transition radiation \cite{Jackson,Ginzburg1945}. Both of these effects have been
implemented in our formalism and Monte Carlo. In Fig. (\ref{fig:image0}) we
show qualitatively the dielectric effect and the transition radiation together
with their characteristic frequency $\omega_{de}$.

\section{Results}
\label{results}
Expression \eqref{quantum_intensity} can be numerically evaluated for
arbitrary interaction models, for finite size and arbitrarily structured
targets with dielectric suppression and transition radiation effects
included. We have developed a Monte Carlo code to evaluate this intensity by
means of discretized paths with a typical step of $\delta z =0.1\lambda$. In a
typical run we computed $10^4$ paths for 50 frequencies and 100 photon angles,
spanning from $\sim$ $10^3$ steps for the shortest medium to $\sim$ $10^5$
steps for the largest. In order to check that our results are correct we
implemented also the Fokker-Planck approximation
\eqref{moliere_solution_gaussian} for \eqref{elastic_weight} in this
discretized approach and compared with the $\delta z\to 0$ limit of
\eqref{quantum_intensity}, which within this approximation produces six
integrable Gaussian path integrals extending the Boltzmann transport approach
\cite{Migdal1956} to finite mediums.
\begin{figure}
\includegraphics[scale=0.6]{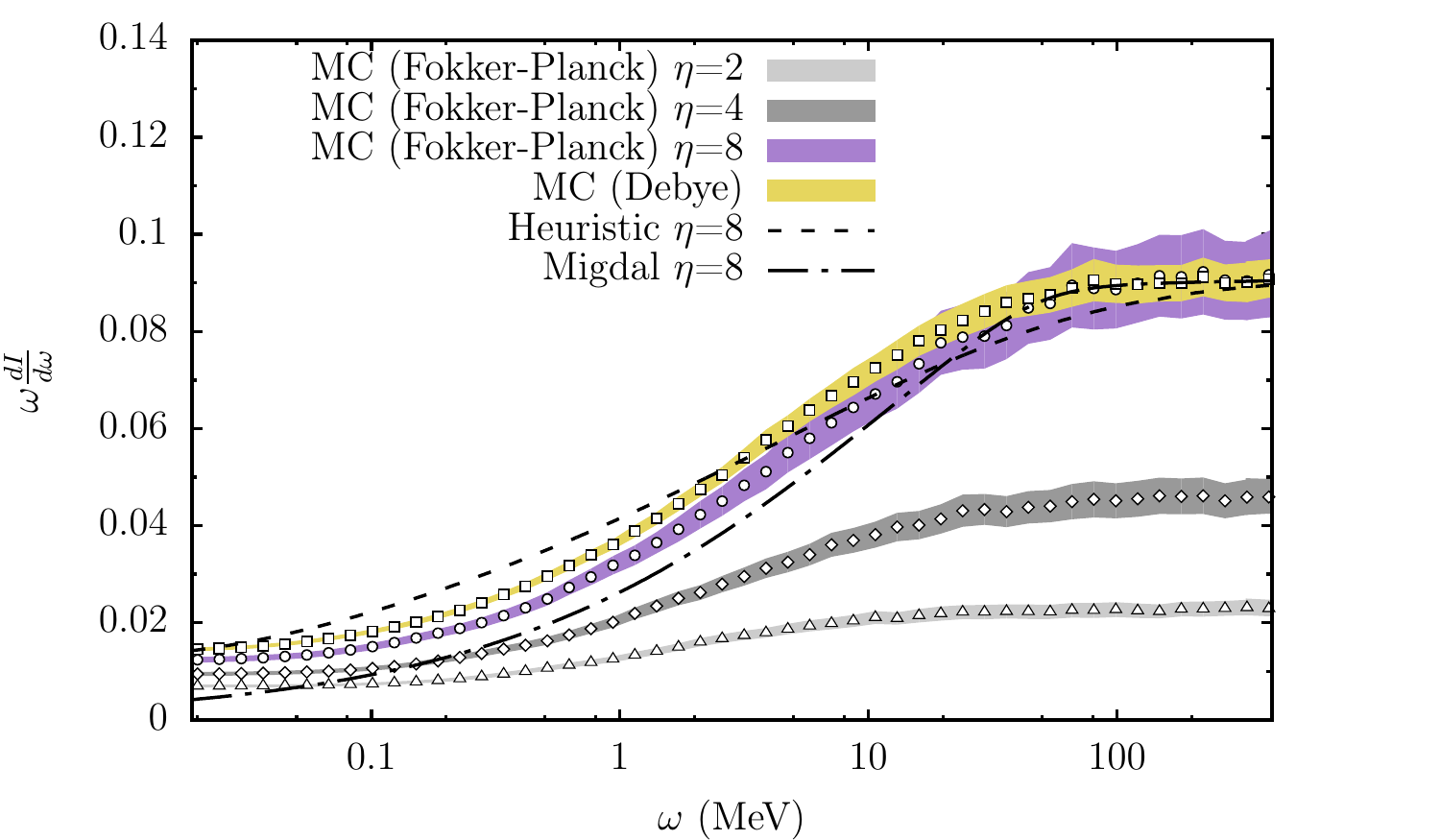}
\includegraphics[scale=0.6]{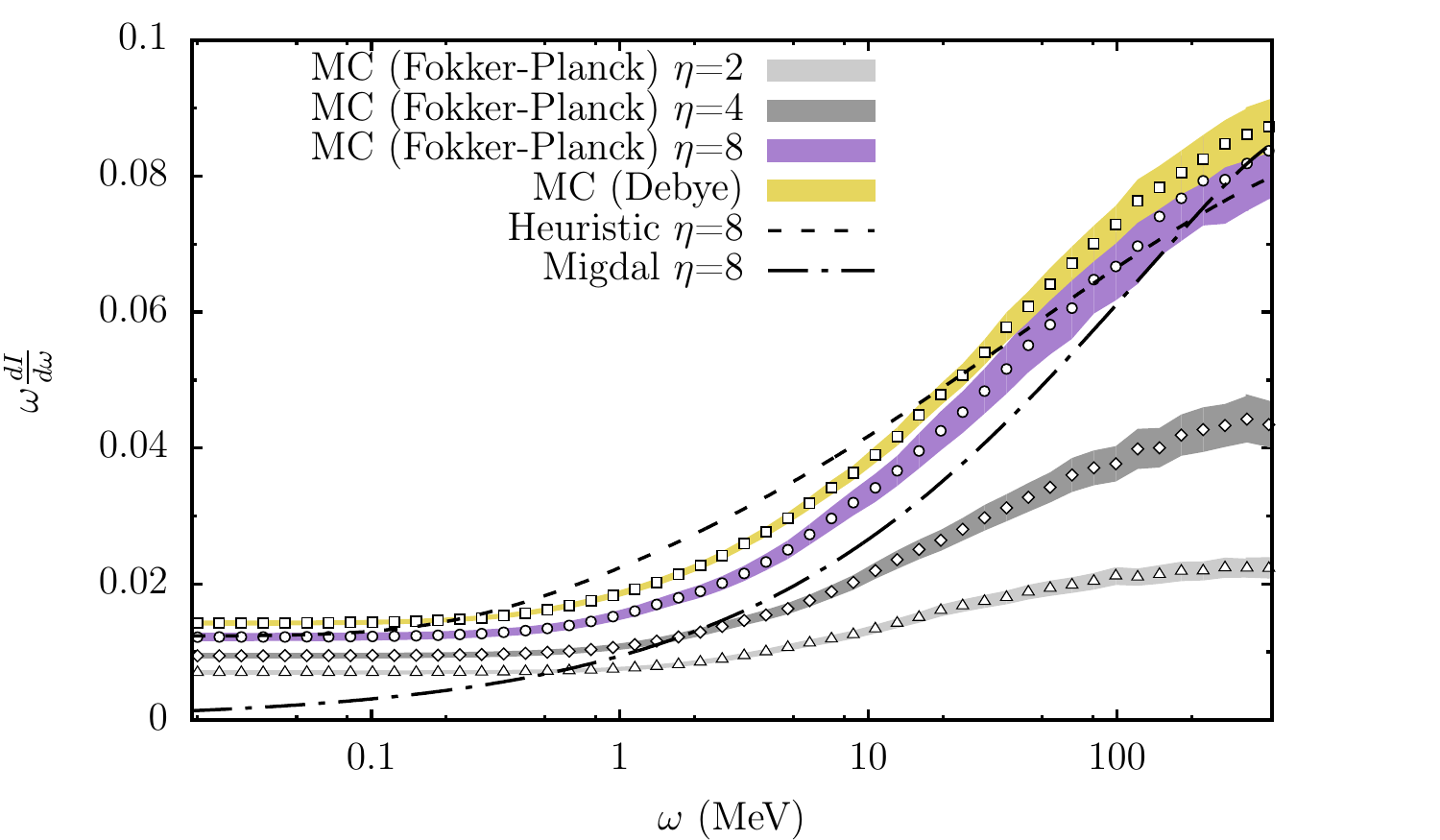}
\caption{Intensity of photons emitted by an electron of $p_0=8$ GeV (top) and
  $p_0=25$ GeV (bottom) after traversing a sheet of gold of length $l=0.2$ mm
  in the Monte Carlo (MC) evaluation of \eqref{quantum_intensity} in the Debye
  interaction (squares), the Fokker-Planck approximation with $\eta=8$
  (circles), $\eta=4$ (diamonds) and $\eta=2$ (triangles). Also shown our
  formula \eqref{heuristic_formula} and the Migdal prediction
  \eqref{migdal_prediction}.}
\label{fig:image3}
\end{figure}
We present our result for the Debye and Fokker-Planck cases for targets of
lengths $l=$ 0.0038, 0.023 and 0.2 mm, corresponding to an average of $n_c =$
142, 862 and 7502 collisions, respectively, for electrons of $p_0 =$ 8 and 25
GeV, in order to compare to the SLAC data presented in
\cite{Anthony3,Anthony1,Anthony2}. A systematic study and comparison with
other experimental results will be presented elsewhere
\cite{FealVazquez_prep}.  For gold we obtain an estimate for the Debye mass of
$\mu_d=16$ keV, a transport parameter of $\hat{q}=(\eta/2)\times 1.89$
keV$^3$, an effective momentum transfer of $\eta=8$ Debye masses in a single
collision and a plasma frequency of $\omega_p=0.080$ keV (see also
\cite{Tsai}).
In Fig.(\ref{fig:image1}) we show the differential photon intensity as a
function of the photon energy for various fixed emission angles for an
electron of $p_0=8$ GeV traversing a gold sheet of $l= 0.023$ mm. The path
integral limit is also shown, and an excellent agreement with the
Fokker-Planck Monte Carlo evaluation is found. At low angles, the
Fokker-Planck approximation overestimates the intensity by $\sim 20\%$.
However, at larger angles the Fokker-Planck approximation underestimates the
intensity, in particular only half of the real emission is obtained for
$\theta=10\gamma^{-1}$.
\begin{figure}
\includegraphics[scale=0.6]{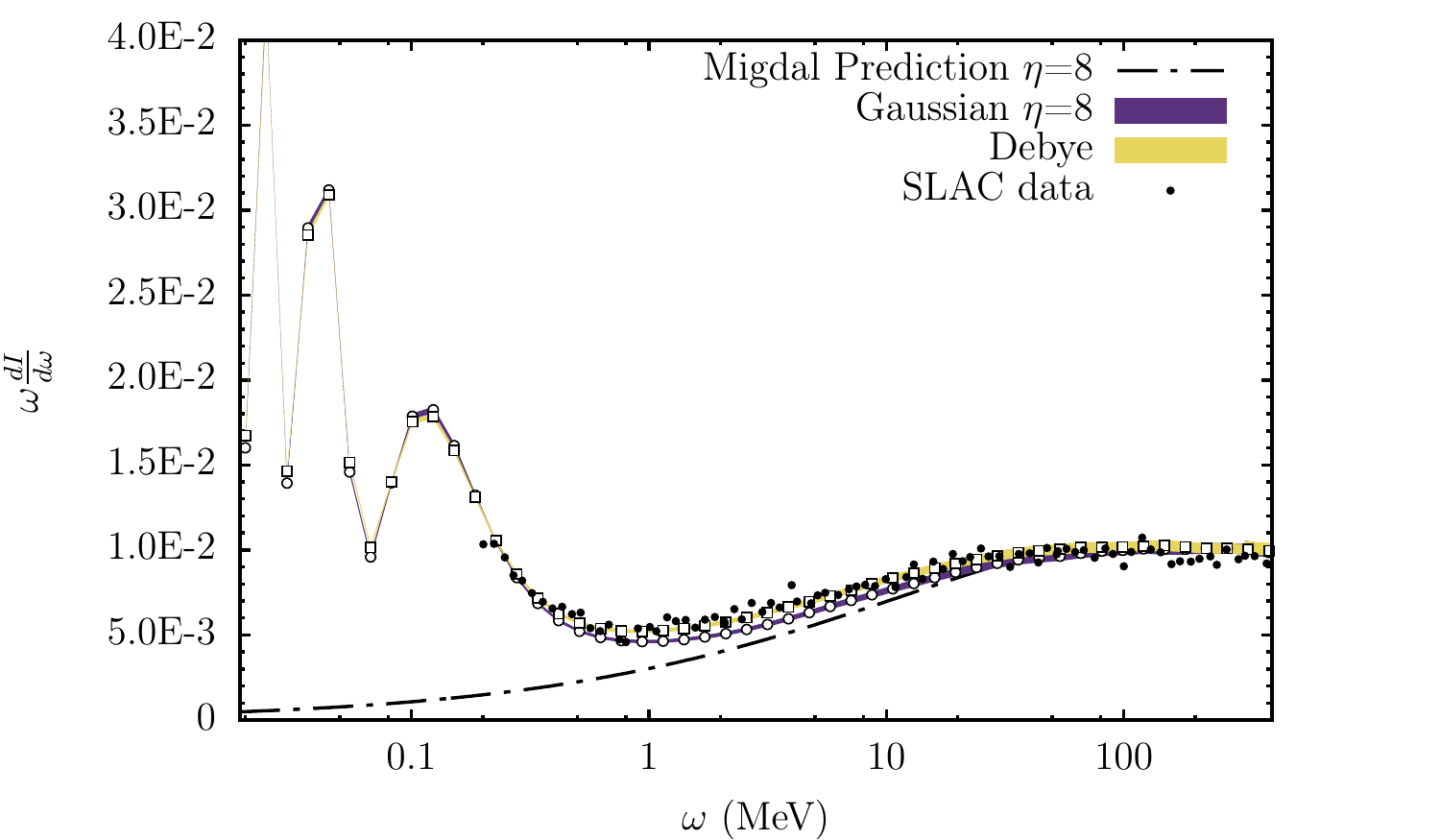}
\includegraphics[scale=0.6]{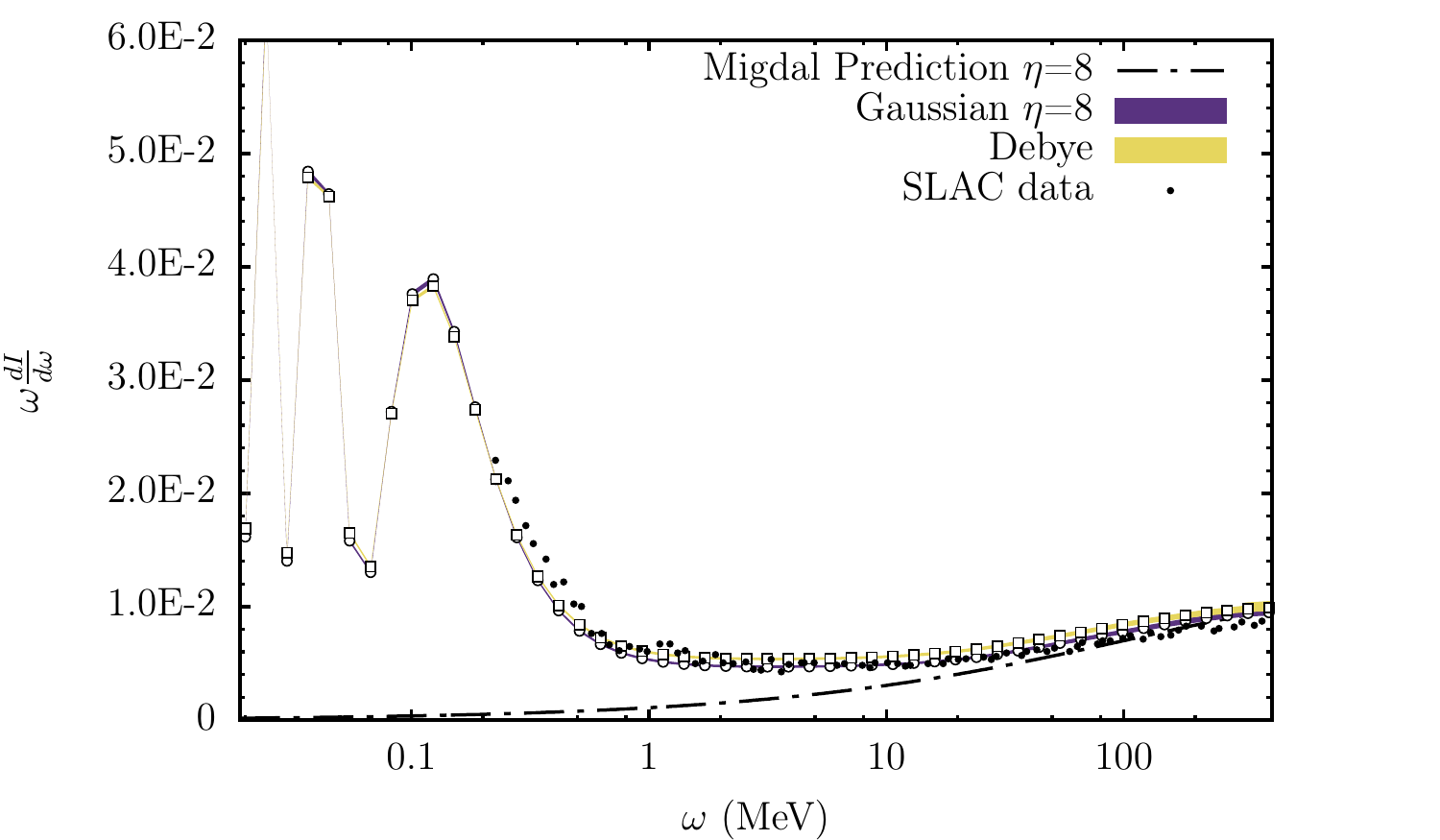}
\caption{Intensity of photons emitted by an electron of $p_0=8$ GeV (top) and
  $p_0=25$ GeV (bottom) after traversing a sheet of gold of length $l=0.023$
  mm in the Debye interaction (open squares) and the Fokker-Planck
  approximation with $\eta=8$ (circles) with the dielectric and transition
  radiation effect included, compared to SLAC experimental data
  \cite{Anthony3} (solid circles) rescaled.}
\label{fig:image4}
\end{figure}
In Fig.(\ref{fig:image2}) we show the angular integrated spectrum for a sheet
of gold of $l=0.0038$ mm for electron energies of $p_0=8$ and 25 GeV.  We see
that a fix of $\eta=8$ Debye masses of momentum transfer at each collision in
the Fokker-Planck approximation matches the incoherent plateau but mismatches
the coherent plateau. Also shown are the expression \eqref{heuristic_formula}
and the Migdal prediction \eqref{migdal_prediction} both in the Fokker-Planck
approximation. The predicted characteristic frequencies are $\omega_c=8$ MeV
and $\omega_s=1.1$ GeV for $p_0=8$ GeV, and $\omega_c=$ 80 MeV and $\omega_s=$
11 GeV for $p_0=$ 25 GeV, being in good agreement with the obtained Monte
Carlo results.

In Fig.(\ref{fig:image3}) we show the same results for a sheet of gold of
$l=0.2$ mm. We see that Migdal prediction becomes a good approximation for
$n_c\ge 10^4$, i.e. when the coherent plateau can be neglected. For this
length we predicted $\omega_c=8$ keV and $\omega_s=60$ MeV for $p_0=8$ GeV,
and $\omega_c=80$ keV and $\omega_s=588$ MeV for $p_0=25$ GeV. All these
values are in well agreement with the Monte Carlo evaluation.

In Fig.(\ref{fig:image4}) we show the dielectric and transition radiation
effect implementation both in the Debye interaction and the Fokker-Planck
approximation, and compare with SLAC data \cite{Anthony3}, for a sheet of gold
of $l=0.023$ mm and electron energies of $p_0=8$ and 25 GeV. The
characteristic frequency predictions $\omega_{de}= 0.6$ MeV for $p_0= 8$ GeV
and $\omega_{de}=1.9$ MeV for $p_0=25$ GeV and the comparison with
experimental data are in very good agreement. The LPM characteristic
frequencies are given in this case by $\omega_c= 0.48$ MeV and $\omega_s= 418$
MeV for $p_0= 8$ GeV, and $\omega_c= 4.7$ MeV and $\omega_s= 4$ GeV for
$p_0= 25$ GeV.
\begin{figure}
\includegraphics[scale=0.6]{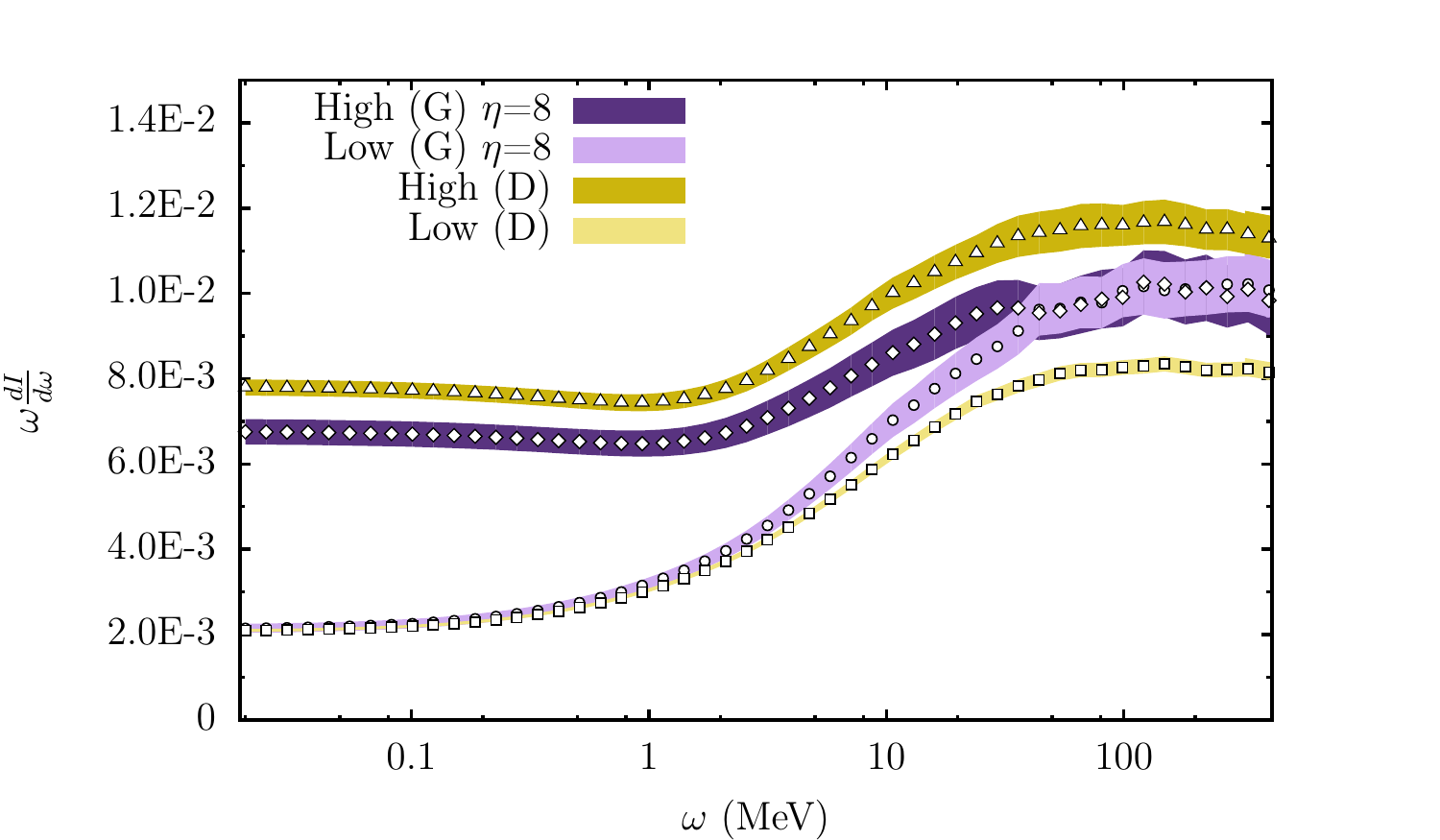}
\caption{Intensity of photons emitted by an electron of $p_0= 8$ GeV after
  traversing a sheet of gold of length $l= 0.023$ mm from electrons with final
  transverse momentum $\v{p}_t>4$ MeV in the Debye interaction (triangles) and
  the Fokker-Planck approximation with $\eta=8$ (diamonds), and with
  $\v{p}_t<4$ MeV in the Debye interaction (squares) and the Fokker-Planck
  approximation with $\eta=8$ (circles).}
\label{fig:image5}
\end{figure}

In Fig.(\ref{fig:image5}) we show the intensity of bremsstrahlung
for electron with a final transverse momentum $|\delta\v{p}|<4$ MeV or 
$|\delta\v{p}|>4$ MeV. We see that the Fokker-Planck results do not reproduce
well the Debye calculation. Although large differences could be expected 
for the case of $|\delta\v{p}|>4$ MeV, as the Fokker-Planck approximation
underestimate the long tails of the tranverse momentum distribution, it is
perhaps more surprising to find that also cutting at low $p_t$ produces
different results: the Fokker-Planck result overestimates the emission in this
case at large frequencies. 

\section{Conclusions}
\label{conclusions}

A formalism implemented with a Monte Carlo method has been presented which is
able to evaluate the bremsstrahlung intensity in a multiple scattering
scenario under a general interaction. We have also found an heuristic formula
which describes the LPM effect for finite size targets. Dielectric and
transition radiation effects related to effective photon masses in the medium
dispersion relation are included in this formalism if needed.  Our results
reproduce the experimental data of SLAC.

We have shown that the Fokker-Planck approximation does not fit well the
differential angular spectrum, especially if kinematical cuts are applied in
the final particles. In the integrated spectrum, the Fokker-Planck
approximation fails to reproduce the spectrum. If the $\hat{q}$ is fixed using
the incoherent plateau, then the coherent plateau is not well reproduced, 
unless a length dependent definition of the transport properties of the medium
is used.

\begin{acknowledgments}
We thank J. Alvarez-Mu\~niz, N. Armesto, C. Salgado, and J. Sanchez-Guillen
for reading the manuscript and for discussions. We thank the grant Mar\'{\i}a
de Maeztu Unit of Excelence of Spain and the support of Xunta de Galicia under
the project ED431C2017. This paper has been partially done under the project
FPA2017-83814-P of MCTU (Spain).
\end{acknowledgments}

\end{document}